\documentclass[a4paper,10pt]{amsart}
\usepackage{xcolor}
\usepackage[utf8]{inputenc}
\usepackage{amsmath}
\usepackage{amssymb}
\usepackage{bbold}
\usepackage{hyperref}

\hypersetup{colorlinks=true,linktoc=page,linkcolor=blue,citecolor=red,urlcolor=cyan}

\title{Worldline Formalism in Snyder Spaces}
\author{S.~A.~Franchino-Vi\~nas}
\address{Theoretisch-Physikalisches Institut, Friedrich Schiller Universit\"at Jena, Max Wien Platz 1, 07743 Jena, Germany.}
\email{sa.franchino@uni-jena.de}

\author{S. Mignemi}
\address{Dipartimento di Matematica e Informatica, Università di Cagliari, viale Merello 92, \\09123 Cagliari, Italy, and \newline
INFN, Sezione di Cagliari, Cittadella Universitaria, 09042 Monserrato, Italy.}
\email{smignemi@unica.it}

\begin{document}

\maketitle

\begin{abstract}
We study the $\phi_{\star}^4$ model for a scalar field in a linearization of the Snyder model, using the methods of the Worldline Formalism. Our main result is a master equation for the 1-loop n-point function. From this we derive the renormalization of the coupling parameters of the theory and observe the appearance of a $\phi^6$ divergent contribution that opens the question of whether this theory is renormalizable or not. Additionally, we observe that some terms in the renormalized action can be interpreted as coming from an effective metric proportional to the square of the field.\\[0.5cm]
\end{abstract}


\section{Introduction}

Many theoretical arguments point at the conclusion that the present understanding
of the structure of spacetime must be modified at short distances if one wants to
reconcile quantum mechanics and general relativity.
In particular, the concepts of point and localizability may not be adequate in this
context, suggesting the possibility of introducing new geometrical concepts in order to
describe spacetime at the Planck scale.

One of the oldest attempts in this sense is the idea of a noncommutative geometry.
The first proposal of this kind was advanced in a paper by Snyder \cite{Snyder}, who conjectured
that the noncommutativity of spacetime coordinates could alleviate the singularities in field theory.
The idea did not raise much interest until more recent times, when noncommutative
geometry has become an important field of research, both in mathematics \cite{Connes} and in
theoretical physics \cite{Doplicher}.

Two classes of models have attracted much attention: the canonical one \cite{Douglas}, inspired by
the Moyal formulation of quantum mechanics \cite{Moyal}, which assumes constant commutators
between the position coordinates, and the $\kappa$-Poincar\'e model \cite{Lukierski}, in which the
commutators of the coordinates form a Lie algebra.

An important tool in the study of these models has been the introduction of the Hopf
algebra formalism  and the definition of a noncommutative star product, that maps the
product of functions of noncommutative coordinates into a product of commutative ones \cite{Majid}.
This allowed the construction of a quantum field theory (QFT) on noncommutative spaces, using the star
product to deform the product of fields.
Several investigations have been carried out in this way, in relation with canonical \cite{Douglas} and
$\kappa$-Poincar\'e QFT \cite{kQFT}, revealing unexpected features. One of the most interesting findings has been the
discovery of the UV/IR mixing \cite{Minwalla}, i.e.\ the appearance of infrared divergences in
the process of renormalization of the ultraviolet ones.

Comparatively little is known about field theory in Snyder space, in spite of the fact that it enjoys the
important property of preserving the Poincar\'e invariance, contrary to other noncommutative models,
where the invariance is deformed or broken.
Snyder spacetime has been investigated from several points of view \cite{Sn}, and some generalizations
have been proposed \cite{Battisti,Mignemi,Meljanac},
but only recently quantum field theory has been examined besides the tree level.
One of the reasons could be the fact that the Hopf algebra of the Snyder model is nonassociative,
so that the star product depends on the order of the multiplication of the fields, complicating
the already involved calculations of other models and opening the possibility of defining several
non-equivalent interaction terms.

After early proposals for a formulation of the tree-level theory \cite{Battisti,Girelli,Mignemi},
the field theory of a scalar field with quartic self-interaction in a generalized Snyder model has been
studied, in an approximation linear in the noncommutativity parameter $\beta$ \cite{You1}. However, an expansion in $\beta$ does not give
the exact ultraviolet behavior of the theory.
In \cite{You2} the investigation has been extended to all orders in $\beta$.
Unfortunately, due to insurmountable algebraic difficulties, it has been possible to
compute only some of the terms appearing in the one-loop two-point function.
The results point at a renormalizable theory, where however the phenomenon of
IR/UV mixing is present, at least for some choices of the interaction term.

On the other hand, one of the techniques that has proven useful in QFT computations is the Worldline Formalism (WF). Since  Feynman's original idea  to express some QFT quantities in terms of path integrals in a first quantization language \cite{Feynman:1950ir}, the Formalism has been applied to several computations, among them in the calculation of  gravitational anomalies \cite{AlvarezGaume:1983ig}, in quantum gravity \cite{Ribeiro:2018pyo},  abelian and nonabelian gauge theories \cite{Schubert:2001he} and on manifolds  with boundaries \cite{Bastianelli:2008vh}.  The strength of this formalism lies in its possibility to handle symmetries and the way they are automatically displayed in the simplified results.

Lately, the WF has also been introduced in the framework of noncommutative QFT \cite{Bonezzi:2012vr}, where different models on the Moyal plane have been studied.
The interesting results achieved in these works, together with the promising properties of Snyder spaces, have motivated us to generalize the application of the WF to the latter.
Actually and as far as we know, in this paper we provide the first example of the application of these methods to noncommutative QFT on spaces different from the Moyal plane.

The organization of this article is as follows. In Section \ref{sec.snyder} we review the definition and basic properties of Snyder spaces. A formulation of a quartic interacting QFT in Snyder space is presented in Section \ref{sec.qft}, formulation which is specially suited to the use of the WF, which is performed  in Section \ref{sec.WF}. The master formula obtained for the $2n$-point functions, cf.\ equation \eqref{eq.master_formula}, is then used to compute the 2-, 4- and 6-point functions in Section \ref{sec.2p}, \ref{sec.4p}  and \ref{sec.6p}   respectively. In these sections, we  proceed also to the renormalization of the parameters involved in each $n$-point function. Later, we discuss this results and state our conclusions in Section \ref{sec.conclusions}. The rather long expressions of the coefficients involved in the computation of our master formula are left to Appendix \ref{app.coefficients_A}, while the presentation of some relevant results on path integrals in phase space are given in Appendix \ref{app.mean_values}. Finally, some intermediate results regarding the computation of the 4-point function are written in Appendix \ref{app.four_point}.

\section{Generalized Snyder spaces}\label{sec.snyder}
The Snyder space was originally introduced in \cite{Snyder} as an example of a discrete spacetime where Lorentz invariance is not broken. The generalized Snyder spaces are defined then as its deformations such that the noncommutative coordinates $\bar{x}_{\mu}$ and $p_{\mu}$ satisfy the following commutation relations \cite{Mignemi,Meljanac}:
\begin{align}
\begin{split}
 [\bar{x}_{\mu}, \bar{x}_{\nu}] &= i\beta M_{\mu\nu} \psi(\beta p^2),\\
 [p_{\mu}, p_{\nu}]&=0,\\
 [p_{\mu}, \bar{x}_{\nu}] &= -i\phi_{\mu\nu}(\beta p^2),
 \end{split}
 \end{align}
while the Lorentz generators $M_{\mu\nu}$ have the same commutation relations as in the usual case, i.e.
\begin{align}
 \begin{split}
[M_{\mu\nu}, M_{\rho\sigma}]&=i (\eta_{\mu\rho}M_{\nu\sigma}-\eta_{\mu\sigma}M_{\nu\rho}+\eta_{\nu\rho}M_{\mu\sigma}-\eta_{\nu\sigma}M_{\mu\rho}),\\
 [M_{\mu\nu}, p_{\lambda}]&=i (\eta_{\mu\lambda} p_{\nu}-\eta_{\lambda\nu} p_{\mu}),\\
 [M_{\mu\nu}, \bar{x}_{\lambda}]&=i (\eta_{\mu\lambda} \bar{x}_{\nu}-\eta_{\lambda\nu} \bar{x}_{\mu}).
 \end{split}
 \end{align}
In these equations we have introduced the parameter $\beta$, usually called the noncommutativity parameter, the metric $\eta_{\mu\nu}$ of Minkowski space and arbitrary functions $\psi$ and $\phi_{\mu\nu}$, constrained only by the fact that the Jacobi identities should still be valid.

It is customary to perform an expansion for small $\beta$, since by heuristic arguments its presumed scale is of order $M_{pl}^{-2}$. Under this hypothesis we may propose a realization of the noncommutative coordinates $\bar{x}_{\mu}$ in terms of $x_{\mu}$, the commutative ones,
\begin{align}
 \bar{x}_{\mu}= x_{\mu}+ \beta (s_1x_{\mu} p^2+s_2 x\cdot p p_{\mu}+ c p_{\mu}) + \cdots,
\end{align}
where $s_1$, $s_2$ and $c$ are arbitrary real parameters \cite{Mignemi,Meljanac}.

As a consequence of this expansion the original commutation relations are fixed to be
\begin{align}\label{eq.commutation_relations}
\begin{split}
 [\bar{x}_{\mu}, \bar{x}_{\nu}] &= i\beta (s_2-2 s_1) M_{\mu\nu} +\cdots,\\
 [p_{\mu}, p_{\nu}]&=0,\\
 [p_{\mu}, \bar{x}_{\nu}] &= -i \left( \eta_{\mu\nu} (1+\beta s_1 p^2)+ \beta s_2 p_{\mu} p_{\nu}\right) +\cdots.
 \end{split}
 \end{align}
The parameter $c$ does not enter in the commutation relations but is necessary in order to obtain a Hermitian operator for $\bar x_\mu$. In particular, in our case this yields $c=-i\left( s_1+\frac{D+1}{2}s_2\right)$.

 At this point one may follow one of two  paths: to work with functions of the noncommutative operators or introduce a star product $\star$ that preserves the commutation relations \eqref{eq.commutation_relations}. Following the second path, it is straightforward to obtain the following definition of the $\star$-product of two plane waves \cite{Mignemi,Meljanac}:
 \begin{align}\label{eq.star}
  e^{i k \cdot x} \star e^{i q\cdot x}= e^{i D(k,q) \cdot x + i G(k,q)},
 \end{align}
where we have introduced the functions
\begin{align}\label{eq.product}
\begin{split}
 D^{\mu}(k,q)&=k^\mu+q^{\mu}+\beta\left[k^{\mu}\left(s_1q^2+\left(s_1+\frac{s_2}{2}\right)k\cdot q\right)+ q^{\mu}s_2\left(k\cdot q +\frac{k^2}{2}\right)\right]+\mathcal{O}(\beta^2),\\
 G(k,q)&=-i\beta\left(s_1+\frac{D+1}{2}s_2\right) k\cdot q+\mathcal{O}(\beta^2).
\end{split}
\end{align}
In particular, one may show that this product is of course noncommutative and under the integral sign it reduces to the usual commutative product, i.e.
\begin{align}
 \int f(x) \star g(x) = \int f(x) g(x).
\end{align}
The latter property is specific of the Hermitian representation of $\bar x_\mu$ \cite{Mignemi,Meljanac}. Moreover, it turns out that the $\star$-product is nonassociative.
However, contrary to other instances of nonassociative $\star$-product of Moyal type that arise in the context of string models with nontrivial $B$-field \cite{Cornalba}, the Jacobi identities are still satisfied in our case, 
as can be explicitly checked,
so that the only problem related to the nonassociativity is the non-uniqueness of the interaction term.

In the following section we will show how to define the Snyder scalar $\phi^4$ field theory and compute the one-loop correction to its effective action.

\section{Linearized Snyder scalar $\phi_{\star}^4$ QFT}\label{sec.qft}

Consider now a scalar field $\varphi$ in a $D$-dimensional Euclidean spacetime whose action contains a quartic interaction term,
\begin{align}\label{eq.action}
 S[\varphi]=\int \frac{1}{2}\partial_{\mu}\varphi \star \partial^{\mu}\varphi + \frac{m^2}{2}\varphi \star \varphi - \frac{\lambda}{4!} \varphi\star(\varphi\star(\varphi\star\varphi)).
\end{align}
As mentioned above, due to the nonassociativity of the $\star$-product, the interaction term may take different forms. We shall discuss this point later.
An alternative but related approach to this model may be found in \cite{You1,You2}.

As stated in the previous section, one can replace the $\star$-product of two functions under the integral sign with the usual product, so that the kinetic part of the action is identical to the usual commutative one. Moreover, the interaction term can be sligthly simplified by removing one of the $\star$-products. The explicit expression for the interaction $S_I$ after a Fourier transform then becomes
\begin{align}\label{eq.SI}
S_I=-\frac{\lambda}{4!} \int \left(\prod_{j=1}^4\frac{d^4 q_{j}}{(2\pi)^D} \right) (2\pi)^D \delta^{(4)}(D_4(q_1, q_2, q_3,q_4) \,g_3(q_1,q_2,q_3,q_4)\, \tilde{\varphi}_1\,\tilde{\varphi}_2\,\tilde{\varphi}_3\,\tilde{\varphi}_4,
\end{align}
where $\tilde{\varphi}_{k}$ is the Fourier transform of the field $\varphi$ evaluated at the momenta $q_k$, and we have absorbed the noncommutative contributions in the functions
\begin{align}
\begin{split}\label{eq.noncommutative}
 D_{4}^{\mu}(q_1,q_2,q_3,q_4):&=q_1^{\mu}+D^{\mu}(q_2, D(q_3,q_4)),\\
 g_3(q_1,q_2,q_3,q_4):&=1+i G(q_2,D(q_3,q_4))+ iG(q_3,q_4).
\end{split}
\end{align}
At this point two differences between \eqref{eq.SI} and the usual commutative case are patent. Firstly, the presence of the $g_3(\cdot)$ function which at order $\beta$ acts as a twist factor. Secondly, the usual momentum conservation is replaced by the conservation of the modified composition of the momenta given by $D_4(\cdot)$.

As next step, we may employ the path integral procedure to quantize the theory,
\begin{align}
 e^{-\Gamma[\phi]} =\int \mathcal{D}\varphi\, e^{-S[\varphi]+\int dx\, J(x) (\varphi(x)-\phi(x))},
\end{align}
where $\Gamma[\phi]$ is the effective action, $\phi$ is the mean (or classical) field and $J$ is a source that should be replaced in terms of the mean field by inverting
\begin{align}
 \phi(x)=\frac{\int \mathcal{D}\varphi\, e^{-S[\varphi]+\int dx\, J(x) \varphi(x)} \varphi(x)}{\int \mathcal{D}\varphi\, e^{-S[\varphi]+\int dx\, J(x) \varphi(x)}}.
\end{align}
Once we perform an expansion of the functional integral around the classical configuration of the field $\phi(x)$ which minimizes the action, we get the one-loop expansion of the effective action
\begin{align}
\Gamma_{1-loop}[\phi] = S[\phi]+\frac{\mu^{-\epsilon}}{2} \text{Tr} \log A,
\end{align}
where $\mu$ is a  quantity with mass dimension introduced to compensate the change in the dimension $D=4-\epsilon$, and $A$ is the operator which has as kernel the second variation of the action
\begin{align}\label{eq.A_operator}
 A f(x)= \int dy \frac{\delta^2 S}{\delta \varphi(x)\delta \varphi(y)}[\phi] f(y).
\end{align}
As it could be foreseen from the nonlocality of the product \eqref{eq.star} and the expression for the action \eqref{eq.SI}, this operator is non-local. Indeed, one of the contributions of its kernel is given by the second variation of the interaction potential
\begin{multline}\label{eq.second_variation}
 \frac{\delta^2 S_I}{\delta \varphi(x)\delta \varphi(y)}[\phi]  =-\frac{\lambda}{2\cdot 4!}\int \left(\prod_{m=1}^{4} \frac{dq_m}{(2\pi)^D}\right) (2\pi)^D \delta^{(4)}(D_4(q_1,q_2,q_3,q_4))\\
\times g_3(q_1,q_2,q_3,q_4)\sum_{\sigma(i,j,k,l)} e^{-i(q_l x + q_k y)} \,\tilde{\phi}_{i} \tilde{\phi}_{j}  ,
\end{multline}
where the sum is performed over all the possible permutations $\sigma(i,j,k,l)$ of the indices $i,j,k,l=1,\dots, 4$.
However, in order to compute the one-loop contribution in the WF, it would be enough to show that this operator can be recast as a local differential operator.

To proceed with our plan, it is useful to simplify the expression in eq.\ \eqref{eq.second_variation} in the following way. First of all, notice that fixing the dependence of $g_3(\cdot)$ and $D_{4}(\cdot)$ on the integration variables and then performing a sum over all the possible permutations  $\sigma(i,j,k,l)$ of the indices in fields and exponentials in expression \eqref{eq.second_variation} is the same as doing the other way around -- viz. fix the indices in the fields and the exponential and then perform the sum over all the indices permutations in the $g_3(\cdot)$ and $D_{4}(\cdot)$ functions to obtain:
\begin{multline}\label{eq.second_variation2}
\frac{\delta^2 S_I}{\delta \varphi_x \delta \varphi_y}
 =-\frac{\lambda}{4!}\frac{1}{2}\int \left(\prod_{m=1}^3 \frac{dq_m}{(2\pi)^D}\right) \tilde{\phi}_{1} \tilde{\phi}_{2} e^{-i (q_4 x+  q_3 y)} \sum_{\sigma(i,j,k,l)} g_3(q_i,q_j,q_k,q_l) \\
\times {\det}^{-1} (\partial_{q_4} D_4(q_i,q_j,q_k,q_l)  )\rvert_{D_{4}(q_i,q_j,q_k,q_l)=0}.
\end{multline}
We have chosen to use the Dirac delta function to perform the $q_4$ integral for reasons that will be soon clear, and it is understood that $q_4$ is to be evaluated at the solution of $D_{4}(q_i,q_j,q_k,q_l)=0$. This evaluation  can be perturbatively performed by considering the linearized expression of eq. \eqref{eq.noncommutative}.

After performing the sum over the permutations we obtain the result
\begin{multline}\label{eq.second_variation_almost}
 \frac{\delta^2 S_I}{\delta \varphi_x \delta \varphi_y} =-\frac{\lambda}{4!}\frac{1}{2}\int \left(\prod_{m=1}^3 \frac{dq_m}{(2\pi)^D}\right) \tilde{\phi}_{1} \tilde{\phi}_{2} e^{-i q_3 y+  i ( q_1+q_2+q_3) x}  \left[4!+ 4\beta (s_1 + s_2)\phantom{\sum_{i=1}^{3}} \right.\\
 \times \left. \left( -2i  \sum_{k}   q_k^2 q_k\cdot x +(2 + D)  \left(- \sum_{i=1}^{3} q_i^2+ q_4^2\right)  \right)\right]_{q_4=-(q_1+q_1+q_3)}.
\end{multline}
Replacing this result in expression \eqref{eq.A_operator} for the $A$ operator, it can be seen that the interaction contribution $A_I$ acts on an arbitrary function $f(x)$ as
\begin{align}\begin{split}\label{eq.A_I}
 A_I f(x) &=-\frac{\lambda}{4!}\frac{1}{2}\int  \frac{dq_1}{(2\pi)^D}\frac{dq_2}{(2\pi)^D} \tilde{\phi}_{1} \tilde{\phi}_{2} e^{ i x ( q_1+q_2)}\\e^{-i(q_l x + q_k y)}
 &\times \left[4!+\beta\left(a_{\mu\nu}(x) (-i\partial^{\mu}) (-i\partial^{\nu})+  b_{\mu}(x) (-i\partial^{\mu})+c(x) \right)\right] f(x),
 \end{split}
 \end{align}
 i.e.\  we have reached our goal of recasting it as a local differential operator.
The expressions for the coefficients $a_{\mu\nu}$, $b_{\mu}$ and $c$, which depend on $x$ but also on $q_1$ and $q_2$, are left to the Appendix \ref{app.coefficients_A}.

\section{Worldline Formalism in Snyder spaces}\label{sec.WF}
Once we have realized that the operator $A$ is nothing but a local differential operator, we can think of it as the Hamiltonian of a fictitious particle in quantum mechanics, with the peculiarity that in this case its potential is momentum dependent \cite{Bonezzi:2012vr}. The trace of $A$ can be consequently computed as a Feynman path integral in phase space, namely
\begin{align}\label{eq.trace_A}
\frac{1}{2} \text{Tr}\log  A 
 =-\frac{1}{2} \int_0^{\infty} \frac{dT}{T}
  \int_{\substack{PBC}} \mathcal{D}p(t) \mathcal{D}x(t) e^{-\int d t [p^2-ip\dot{x}+m^2-V_W(x,p)]},
\end{align}
where $PBC$ means that the integral should be performed over paths $x(t)$ that satisfy periodic boundary conditions, and $V_W$ is the Weyl-ordered kernel of the $A_I$ operator defined in eq.\ \eqref{eq.A_I}, where the  derivatives $(-i\partial_{\mu})$ have been replaced by momentum operators in a first quantization ($p_{\mu}$).

It is important to notice that one must use the Weyl-ordered potential $V_W$ in order for expression \eqref{eq.trace_A} to be valid. In general terms, it means that we should write the potential in a symmetrized way on the variables $x$ and $p$, adding the needed terms coming from the  commutations performed to reach the symmetrization\footnote{ As a simple example consider the Weyl-ordered expression for the product $
 \left(x p\right)_W= \frac{1}{2}(p x+ x p)+\frac{i}{2} $. A more detailed treatment of this issue can be found in \cite{Bastianelli:2006rx}.}.

Turning back to eq. \eqref{eq.trace_A} for the trace of $A$, the Weyl-ordered potential can be cast as
\begin{align}\label{eq.vW}
 V_{W}=\frac{\lambda}{4!}\frac{1}{2}\int  \frac{dq_1 dq_2}{(2\pi)^{2D}} \left[4! e^{ i x ( q_1+q_2)} +\beta (\alpha_{\mu\nu} p^{\mu} p^{\nu}+\beta_{\mu} p^{\mu}+\gamma)\right] \tilde{\phi}_{1}\tilde{\phi_2},
\end{align}
where the exact expressions for the coefficients of this potential are written in Appendix \ref{app.coefficients_A}. It is worth to mention that we will introduce the primed coefficients $\alpha'$, $\beta'$ and $\gamma'$, which correspond to the removal of the $x$ dependence in the non-primed coefficients by performing an integration by parts --  their expression can also be found in Appendix \ref{app.coefficients_A}. The use of both primed and non-primed coefficients has  some advantages, as we will see.

Another comment about equation \eqref{eq.trace_A} is still in order. The usual procedure would be to introduce the mean values
\begin{align}\label{eq.mean_value}
 \left\langle f(x,p)\right\rangle_{PBC} =\frac{\int_{PBC}\mathcal{D}q\mathcal{D}p\,e^{-\int_0^{1} dt\,\left(p^2-ip\,\dot{q}\right)}
f(x,p)}{\int_{PBC}\mathcal{D}q\mathcal{D}p\,e^{-\int_0^{1} dt\,\left(p^2-ip\,\dot{q}\right)}},
\end{align}
which may be easily calculated after the computation of   the generating functional $Z_{PBC}[k,j]$, with sources $k$ and $j$ corresponding to the paths $p$ and $x$ respectively. However, the kinetic operator defined on functions that satisfy periodic boundary conditions cannot be inverted, since there exits a zero mode, the constant path $x=x_0$, $p=0$. Our choice is to include an explicit integration over this zero mode and perform the path integrals over paths that satisfy  Dirichlet boundary conditions (DBC). To this end we will use the explicit notation $x(t)=q(t)+x_0$, to write every path $x(t)$ satisfying PBC, in terms of a path $q(t)$ satisfying DBC. Once more, the computation of mean values defined for DBC may be performed using the generating functional $Z_{DBC}[k,j]$, whose expression may be found in Appendix \ref{app.mean_values}.

Using equation \eqref{eq.vW} and the notation of mean values, the calculation of the one-loop $2n$-point function is straightforward -- we just need to expand the exponential of the potential $V_W$ in expression \eqref{eq.trace_A} and collect all the terms which contain a product of a number $2n$ of  $\phi$ fields. The expression for the $2n$-point function obtained from \eqref{eq.trace_A} is then the following
 \begin{align}\label{eq.2n_point_pre}
\begin{split}\Gamma^{(2n)}_{1-loop}&= -\frac{\mu^{-\epsilon}}{4! (n-1)!}\frac{\lambda^n}{2^{n+1}}\int_0^{\infty} \!\!\!\frac{dT}{\left(4\pi \right)^{D/2}}   \frac{e^{-m^2 T}}{T^{D/2+1-n}} \int_{-\infty}^{\infty} dx_0 \int \sideset{}{'}  \prod_{j,k,i}  dt_j dp_{k} \, \psi_i \\
\times &\left\langle e^{i \sum_{i=1}^{n-1} p_i x(t_i)+i(q_1+q_2) x(t_0)}\left[\frac{4!}{n}+\beta  \left(\frac{1}{T} \alpha'_{\mu\nu} p^{\mu}(t_0)p^{\nu}(t_0) +\gamma' \right) \right]\right\rangle_{DBC}  ,
 \end{split}\end{align}
where in order to keep the readability of this expression we have introduced the compact notation for the integrals
 \begin{align}
 \int \sideset{}{'}\prod_{j,k,i}  dt_j dp_{k} \psi_i= \int_0^1\prod_{j=0}^{n-1} dt_j  \int_{-\infty}^{\infty}\prod_{k=1}^{n-1} \frac{dp_{k}}{(2\pi)^D} \frac{dq_{1}}{(2\pi)^D}\frac{dq_{2}}{(2\pi)^D} \prod_{i=1}^{n-1}\psi_i,
\end{align}
 and the $\psi$ functions, the Fourier transform of the  square of the field $\phi$,
 \begin{align}
  \mathcal{F}(\phi^2)(q_i) = \psi_i .
 \end{align}
Now by using the results stated in Appendix \ref{app.mean_values} regarding the generating function with DBC $Z_{DBC}$, the mean value in eq.\ \eqref{eq.2n_point_pre} can be computed to readily obtain our master formula for the $2n$-point function of the effective action,
 \begin{align}\label{eq.master_formula}
\begin{split}\Gamma^{(2n)}_{1-loop}&=-\frac{\mu^{-\epsilon}}{4! (n-1)!}\frac{\lambda^n}{2^{n+1}}\!\!\int_0^{\infty} \!\!\!\!\frac{dT}{\left(4\pi \right)^{D/2}}  \frac{ e^{-m^2 T}}{T^{D/2+1-n}}\int \sideset{}{'}  \prod_{j,k,i}  dt_j dp_{k} \, \psi_i \\
&  \hspace{-0.5cm} \delta^{(4)}(P) \ e^{-\frac{T}{2} (J, B_{22}^{-1} J) } \left\lbrace \frac{4!}{n}+ \beta \left[ \frac{1}{T} \alpha'_{\mu\nu} \left( B^{-1}_{11} \delta^{\mu\nu}- \frac{T}{4}  C[J^{\mu}] C[J^{\nu}] \right) +\gamma'  \right] \right\rbrace ,
 \end{split}\end{align}
in terms of the total momentum $P$, the source $J$ and the $C$ functional, defined as
 \begin{align}
 \begin{split}
  P&=\sum_{j=1}^{n-1} p_j +q_1+q_2,\\
  J&=\sum_{i=1}^{n-1} \delta(t-t_i) p_i+\delta(t-t_0) (q_1+q_2),\\
   C[f]&=\int_0^1 dt \left( B^{-1}_{12}(t_0,t)+B^{-1}_{21}(t, t_0) \right) f(t).
 \end{split}
 \end{align}

Some remarks are now in order. First of all, in spite of the noncommutativeness, the integral over the zero mode guarantees the conservation of the momentum for the $n$-point function.

Moreover, the expansion \eqref{eq.master_formula} has many points of coincidence with what is called the small propertime expansion of the Heat-Kernel of the operator $A$ \cite{Vassilevich:2003xt}. In the latter, one is usually interested in the small $T$ expansion of the exponential operator $e^{-T A}$. Additionally, in such an expansion the coefficients are given by powers of the potential, its derivatives and invariant quantities obtained from the metric. In eq.\ \eqref{eq.master_formula}, the $\beta$-independent factor gives the usual commutative contribution to the effective action in the case of a matrix-valued potential, with the integral over the temporal $t$ variables reproducing the factors  obtained in the literature for the small $T$ expansion.

It should be noted however, that  noncommutativeness introduces some new features. Indeed, the noncommutative part contains two contributions:
the first one is the term  proportional to $\alpha'$. In the context of the ``regular" Heat-Kernel technique such a contribution wouldn't be expected to arise, because  the potential factor is itself proportional to the proper time. There exists nevertheless one possibility: that this term could be interpreted as coming from the small $\beta$ expansion of a   metric, which in our case should therefore be proportional to the second power of the field $\phi$. This  would imply that the $O(T^{-1})$ term has its origin in the expansion of the  invariant measure $\sqrt{g}$ factor, while the $O(T^0)$ should be the analogue of the curvature term arising in the study of a scalar field on a curved background. This is consistent with  the fact that this contribution arises from the $p^2$ factor of the noncommutative potential $V_W$, very much akin to the general expression for a path integral in curved spaces\footnote{The derivation of this formula may be found in \cite{Bastianelli:2006rx}. In this formula $g^{ij}$ is the inverse metric, $R$ the curvature scalar, $\Gamma^{j}_{kl}$ the Christoffel symbols of the metric and Einstein's convention for summation is used. }
\begin{align}
\int \mathcal{D}x \mathcal{D}p\, e^{-\int dt p_i g^{ij}(x) p_j -i p x+\frac{1}{8}(R+g^{ij} \Gamma^{l}_{ik}\Gamma^{k}_{jl}) +V(x)}.
\end{align}

Of course this claim is only valid for our first-order expansion in $\beta$ and an all-order generalization cannot be immediately stated. In effect, were our expansion valid for any power of $\beta$, the $n$-point function would have shown contributions that have negatives powers in $T$ for any $n$ and could spoil the renormalizability of the model.

The second contribution comes from the $\gamma'$ factor in \eqref{eq.master_formula}. This is the usually called potential contribution.

After having discussed these general aspects of the $n$-point function, we now turn our attention to the study of the renormalization of our $\beta$-linearized theory. As can be seen from eq.\ \eqref{eq.master_formula} this  is only needed for  the $2$-, $4$-and $6$-point function.

\section{Two-point function of the linearized $\varphi_{\star}^4$ theory}\label{sec.2p}
Let us consider the one-loop contribution to the two-point function, which in the notation of eq.\ \eqref{eq.mean_value} corresponds to the mean value of the potential:
\begin{align}
 \Gamma^{(2)}_{1-loop}&=-\frac{ \mu^{-\epsilon}}{2} \int_0^{\infty} \frac{dT}{\left(4\pi T\right)^{D/2}} e^{-m^2 T}\int_{-\infty}^{\infty} \!\!\!\!\!dx_0 \left\langle\int_0^1 d\tau_1 V_W(\sqrt{T} q(\tau_1)+ x_0, r(\tau_1))\right\rangle_{DBC}.
 \end{align}
The expression gets further simplified by performing the $x_0$ integral. Indeed, it gives a delta function that one can use to compute the $q_2$ integral.
Using the results of the previous section to compute the mean expectation values, we obtain the expression
 \begin{align}
 \Gamma^{(2)}_{1-loop}
 &=-\frac{\lambda \, \mu^{-\epsilon}}{4\cdot 4!}\frac{ m^{D-2} }{(4\pi)^{D/2}}\left[ 4! + 8(s_1+s_2) (D+2) \beta  m^{2} \right]\,\Gamma\left(1-\frac{D}{2}\right) \int dx \,\phi^2(x).
 \end{align}
According to this result, in analogy with the commutative case, we should renormalize the mass but not the field. The renormalization process in the minimal subtraction (MS) prescription in $D=4-\epsilon$ dimensions is simply as follows:
\begin{align}\label{eq.renormalization_twopoint}
\begin{split}
 m^2&=m_0^2\left[1 +\frac{\lambda_0}{(4\pi)^2} \left( 1+2(s_1+s_2) \beta m_0^{2} \right) \frac{ \mu^{-\epsilon}}{\epsilon} \right],
\end{split}
\end{align}
in terms of the bare parameters $m_0$ and $\lambda_0$. From \eqref{eq.renormalization_twopoint} the mass beta function can be readily obtained\footnote{In order to avoid confusion with the noncommutative parameter $\beta$, we will always write the beta function of a given coupling $x$ as $\beta_{x}$.     }:
\begin{align}
\begin{split}
 \beta_{m^2}&:=\frac{\partial m^2}{\partial \log \mu}\\
 &=-\frac{\lambda}{16\pi^2} \left(1+2  (s_1+s_2)  \beta m^{2} \right) m^2.
 \end{split}
\end{align}

\section{Four-point function of the linearized $\varphi_{\star}^4$ theory}\label{sec.4p}
We may also employ eq.\ \eqref{eq.master_formula} to analyze the renormalization of the four-point function. Restricting to the terms giving the divergent contributions and after performing the proper time integral, we get
\begin{align}\label{eq.four_point2}
\begin{split}
\Gamma^{(4)}_{1-loop}&=-\frac{1}{4! }\frac{\lambda^2}{2^3} \frac{\mu^{-\epsilon}}{(4\pi )^{D/2}} \int dx \left\lbrace \phantom{\frac{1}{2}}\hspace{-0.4cm} -2 \beta  (s_1+s_2) D(D+2) m^{D-2} \Gamma(1 - D/2) \phi^4 \right.\\
&\hspace{-0.6cm}\left.+ \frac{\Gamma(2 - D/2)}{m^{4-D}} \left[ \frac{4!}{2} \phi^4+ \beta \left( 12 \phi_{\star,(1)}^4 -\frac{4}{9} (s_1+s_2) (D-4) (D+2) \phi^3 \partial^2\phi\right) \right]  \right\rbrace +\text{f.t.},
\end{split}\end{align}
where we have denoted $\phi_{\star,(1)}$ the linear contribution in $\beta$ of the noncommutative quartic interaction under the integral sign and ``f.t.'' means finite terms. Some intermediate steps in the derivation of this formula are left to Appendix \ref{app.four_point}.

The $\beta$-independent contribution in the RHS of  \eqref{eq.four_point2} corresponds clearly to the usual commutative contribution. On the other hand, the terms that we call ``metric'' terms, i.e.\ those depending on $B^{-1}_{11}$ and $C[\cdot]$, sum up with the potential term to give rise to three kind of contributions: one proportional to the commutative quartic interaction, another proportional to $\phi_{\star,(1)}$  and the last proportional to a new nonlocal quartic interaction.

Surprisingly, the last term has an additional factor that renders it finite in the limit $D=4$. Therefore, once we expand these formulae around $D=4-\epsilon$, we see that the renormalization proceeds by introducing appropriate counterterms which have the structure of the original action, if one considers the quartic potential split into the $\beta$-independent and $\beta$-linear contributions. Indeed, the divergent contributions in \eqref{eq.four_point2} read
\begin{align}\label{eq.four_point_div}
\Gamma^{(4)}_{1-loop}=-\frac{\lambda^2}{128 \pi^2}& \frac{\mu^{-\epsilon}}{\epsilon} \int dx\, \left\lbrace \left[ 1 + 4   (s_1+s_2) \, \beta m^{2}  \right] \phi^4 + \beta \phi_{\star,(1)}^4 \right\rbrace + \mathcal{O}(\epsilon^0).
\end{align}
In our linear noncommutative expansion we could either introduce a new coupling constant for the $\phi_{\star,(1)}^4$ term or just interpret eq.\ \eqref{eq.four_point_div} as showing the necessity to proceed to the renormalization of the noncommutative parameter $\beta$. The latter procedure was proved to be required for example in Moyal noncommutative SU(N) gauge theories \cite{Latas:2007eu} in order to save their perturbative renormalization properties.  Using this as motivation we can read the renormalization of the coupling constant $\lambda$ and of the noncommutative parameter $\beta$:
\begin{align}
\begin{split}
\lambda&=\lambda_0 \left[1+ \frac{ \mu^{-\epsilon}}{\epsilon}\frac{3\lambda_0 }{16 \pi^2} \left[ 1 + 4   (s_1+s_2) \, \beta_0 m_0^{2}  \right] \right], \\
 \beta &= \beta_0 \left[ 1- \frac{ \mu^{-\epsilon}}{\epsilon}  \frac{3 \lambda_0}{4\pi^2} \beta_0m_0^2\right].
\end{split}
 \end{align}

The corresponding beta functions are straightforwardly obtained and are
\begin{align}
\begin{split}
\beta_{\lambda}&= -\frac{3\lambda^2}{16 \pi^2}  \left[ 1 + 4   (s_1+s_2) \, \beta m^{2}  \right], \\
 \beta_{\beta} &= \frac{3 \lambda}{4 \pi^2} \beta^2 m^2.
\end{split}
\end{align}
The beta function of the noncommutative parameter shows that the theory is ``asymptotically commutative'' for $\lambda<0$ in the UV, i.e.\ it has a vanishing $\beta$ in this regime. 

\section{Six-point function  of the linearized $\varphi_{\star}^4$ theory}\label{sec.6p}

The only divergent expression left in the effective action is the six-point function. The relevant term can be readily extracted from expression \eqref{eq.master_formula} and is
\begin{align}
\begin{split}
 \Gamma^{(6)}_{1-loop}&= -\frac{\beta}{ 4! } \frac{\mu^{-\epsilon}}{(2\pi)^{3D}}\frac{\lambda^3}{64}  \int_0^{\infty} \frac{dT}{\left(4\pi \right)^{D/2}}   \frac{e^{-m^2 T}}{T^{D/2-1}}\\
&\hspace{1cm} \times\int_{-\infty}^{\infty} dp_1dp_2dq_1dq_2 \,\delta^{(4)}(p_1+p_2+q_1+q_2)\, \psi_1 \psi_{2} \,  \delta^{\mu\nu}\alpha'_{\mu\nu} +\text{f.t.}\,.
\end{split}
\end{align}

Although the $\alpha'$ coefficient contains derivatives acting on the field $\phi$, it can be shown that all the contributions add up to a usual commutative $\phi^6$ interaction that is however divergent as $\epsilon$ tends to zero, for $D=4-\epsilon$:
\begin{align}\label{eq.six_point}
 \begin{split}\Gamma^{(6)}_{1-loop}
 &=-\frac{\beta}{ 4!}\frac{\mu^{-\epsilon}}{(4\pi )^{D/2}}\frac{\lambda^3}{64 } (s_1+s_2)    \left(D -\frac{4}{6}\right) (D+2) \Gamma\left(2-\frac{D}{2}\right) m^{D-4} \!\!\int \phi^6 dz+\text{f.t.}\\
 &=-\frac{5}{4!\,\, 128 \pi^2}  \frac{\beta \lambda^3}{\epsilon} (s_1+s_2) \int \phi^6 dz+\mathcal{O}(\epsilon^0).
\end{split}\end{align}

In order to proceed to the renormalization we will need to introduce one additional local term to the original action, namely a sixth interaction, whose coupling constant would absorb the divergence present in formula \eqref{eq.six_point}. This would create a domino effect in the renormalization procedure. In fact, it can be seen that after the introduction of a $\phi^6$ interaction term in the original action, the presence of the $\alpha'$ term in \eqref{eq.master_formula} implies the creation of a new divergent interaction contribution with an eighth power of the field. Unfortunately this would also force the inclusion of an interaction term with a power ten and so forth, unless a fortuitous combination of the parameters enforces the end of this domino effect.

However, one may also suggest to work with parameters $s_1$ and $s_2$ such that their sum cancels \cite{You1}, in which case this divergent term vanishes. The drawback of this option, is that the linearized theory then reduces to the commutative one.

\section{Conclusions}\label{sec.conclusions}
Our investigation of the linearization in the noncommutative parameter $\beta$
of the Snyder $\phi_\star^4$ Quantum Field Theory (QFT), in the framework of
the Worldline Formalism (WF), has lead us to the calculation of the master equation \eqref{eq.master_formula}, i.e.\
a closed expression for the 1-loop $n$-point functions.
It is worth to notice, that this is the first time calculations on a noncommutative space
different from the Moyal plane are performed using the WF.

In this respect, we find it suggestive the fact that we can interpret the presence of
some noncommutative corrections as due to the existence of an effective metric that explicitly
depends on the mean field $\phi$. This provides a hint towards the heuristically claimed strong interplay between fields and gravity
expected to be found in noncommutative QFTs.

Our results for the 2- 4- and 6-point functions,
 the only ones that need to be renormalized, are in accord with those
obtained in \cite{You1} using different methods. The renormalization of the coupling constants
has then been performed in the MS prescription, with the notable  fact that, in the linearized theory and
up to the one-loop order, the renormalization of the 4-point function could be understood as involving the
renormalization of the
noncommutative parameter $\beta$. Using this interpretation, we find that the theory is ``asymptotically commutative'' for positive $\beta$ and negative $\lambda$, i.e.\ the noncommutative parameter decreases as the energy scale increaeses. Since this behaviour has been also observed in Moyal noncommutative theories\footnote{In the case of SU(N) theories it has been shown that they are also asymptotically free, i.e.\ the coupling parameter decreases as the energy scale decreases, cf. \cite{Martin:1999aq} for the SU(1) model.} \cite{Latas:2007eu}, we feel tempted to ask ourselves  whether this is a universal property of noncommutative theories regardless of
the specific choice of the underlying noncomutative space, even if answering this question is out of the scope of this work.

One of the main outcomes is that the 6-point function gives rise to divergences that
can lead to perturbative non-renormalizability, because the addition of a $\phi^6$
term on the original action would generate a domino effect,
in the sense that terms with arbitrary high powers of the field should also be added.
An exception occurs when the parameters of the theory
obey the relation $s_1+s_2=0$. Curiously in this case, in spite of the
noncommutativity, the $\beta$-linearized one-loop QFT is identical to the
commutative one, because the corrections to the interaction term vanish, cf.\ \eqref{eq.second_variation_almost}.
We are not able to give a physical interpretation of this special relation
between the two parameters. It would be interesting to study such models to
higher orders in $\beta$ to see if this property still holds and
how the renormalizability of the theory is affected.
In any case, it must be recalled that the ultraviolet behavior of the full theory is different
from its first-order expansion in $\beta$ which is studied in this paper, and it is likely that the full theory 
be UV renormalizable \cite{You2}, although a complete proof of this is still lacking.

We also notice that because of the nonassociativity of the star product,
one may choose some $\phi^4$ interaction terms a priori not equivalent to eq.\ \eqref{eq.action}. It can be
checked inside the WF, however, that in the linearized theory they give rise to the same
results as those obtained with the ordering in \eqref{eq.action}, as already noticed in \cite{You1}.

Unfortunately, at this level it is not possible to discuss the occurrence of
UV/IR mixing, which is one of the most interesting effects associated to
noncommutative QFT. In \cite{You2}, it has been shown that this effect may occur
in the full theory for some choice of the ordering in the potential. However, the terms that lead to
the UV/IR mixing vanish at
the linearized level, so that it is not possible to establish from the present calculations whether this
effect takes place or not. A higher order computation in the $\beta$ parameter is currently being considered.

As a final remark, we notice that in spite of the fact that the quantum mechanics of noninteracting particles
in Snyder spacetime has been claimed to be trivial (i.e.\ equivalent to that on Minkowski spacetime \cite{AA}),
the interacting theories, as our QFT, turn out to be highly nontrivial.

\bigskip

\noindent\textbf{Acknowledgements}:  This work was partially supported by GESTA - Fondazione di Sardegna. SAF acknowledges support from  the DAAD and the Ministerio de Educación Argentino under the ALE-ARG program. SAF would like to thank the Università di Cagliari and specially  SM for their  hospitality. The authors thank J. Trampeti\'c for his useful comments.

\appendix
\section{Coefficients}\label{app.coefficients_A}
The $a_{\mu\nu}$, $b_\mu$ and $c$ coefficients introduced in eq. \eqref{eq.A_I},  Section \ref{sec.qft}, to define the $A$ operator are polynomials in the position $x$ and the momenta $q_1$ and $q_2$, and their explicit form reads
\begin{align}\label{eq.latin_coefficients}
\begin{split}
a_{\mu\nu}(x)&=8i (s_1+s_2) \left(2x^{\mu} (q_1+q_2)^{\nu}+ (q_1+q_2)\cdot x \delta_{\mu\nu}\right),\\
b_{\mu}(x)&= 8i (s_1+s_2)\left( x_{\mu} (q_1+q_2)^2+2(q_1+q_2)\cdot x\, (q_1+q_2)_{\mu}\right)\\
 &\hspace{6cm}+ 8 (2 + D) (s_1 + s_2) (q_1 + q_2)_{\mu},\\
c(x)&=8i(s_1+s_2) \left( (q_1\cdot x)(2q_1\cdot q_2+q_2^2)+(q_2\cdot x)(2q_1\cdot q_2+q_1^2) \right)\\
 &\hspace{6cm}+ 8 (2 + D) (s_1 + s_2)  q_1\cdot   q_2.
\end{split}
\end{align}

On the other side, the Weyl-ordered formula \eqref{eq.vW} involves the coefficients $\alpha_{\mu\nu}$, $\beta_{\mu}$ and $\gamma$, which can be expressed in terms of those in \eqref{eq.latin_coefficients}:
\begin{align}
 \begin{split}
 \alpha_{\mu\nu}&=a_{\mu\nu} e^{ i x ( q_1+q_2)},\\
 \beta_{\mu}&=b_{\mu}+\frac{i}{2} \partial^{\mu}\left( (a_{\mu\nu}+a_{\nu\mu})e^{i (q_1+q_2) x} \right),\\
 \gamma&=c-\frac{1}{4}\partial^{\mu}\partial^{\nu} \left(a_{\mu\nu} e^{i (q_1+q_2) x}\right) +\frac{i}{2} \partial^{\mu} \left(b_{\mu} e^{i (q_1+q_2) x} \right).
 \end{split}
 \end{align}
 From them a  straightforward computation gives the following result:
 \begin{align}\label{eq.alpha_coefficients}
 \begin{split}
  \alpha_{\mu\nu}(x)&=8i (s_1+s_2) \left(2x^{\mu} (q_1+q_2)^{\nu}+ (q_1+q_2)\cdot x \delta_{\mu\nu}\right) e^{ i x ( q_1+q_2)},\\
 \beta_{\mu}(x)&= 0,\\
 \gamma(x)&=4(s_1+s_2) \left( \phantom{\frac{2}{4}} \hspace{-0.35cm}2(q_1\cdot ix)(2q_1\cdot q_2+q_2^2)+2(q_2\cdot ix)(2q_1\cdot q_2+q_1^2) \right.\\
 &\left.\phantom{\frac{2}{4}}\hspace{-0.5cm}-\frac{3}{2} ix\cdot(q_1+q_2)(q_1+q_2)^2- (2 + D)  (q_1^2  + q_2^2) \right) e^{ i x ( q_1+q_2)} .
 \end{split}\end{align}

After performing an integration by parts to cancel the $x$ dependence in the coefficients listed in \eqref{eq.alpha_coefficients} and therefore simplify the computation of the $n$-point function, we get the $\alpha'_{\mu\nu}$, $\beta'_{\mu}$ and $\gamma'$ coefficients:
\begin{align}
\begin{split}
 \alpha'_{\mu\nu}&=-8 (s_1+s_2) \left(2 (q_1+q_2)^{\nu}\partial_{q_1^{\mu}}+ (q_1+q_2)\cdot \partial_{q_1} \delta_{\mu\nu}+ (D+2)\delta_{\mu\nu}\right) \tilde{\phi}_{1}\tilde{\phi}_{2},\\
 \beta'_{\mu}&=0,\\
 \gamma'&=-2(s_1+s_2)  \Big[ 4(2q_1\cdot q_2+q_2^2)(q_1\cdot \partial_{q_1})+4(2q_1\cdot q_2+q_1^2)(q_2\cdot \partial_{q_1})\\
 &\hspace{2cm} -3 (q_1+q_2)^2 (q_1+q_2)\cdot \partial_{q_1}-  (2 + D) (q_1^2-2 q_1 q_2 - 3  q_2^2 ) \Big] \tilde{\phi}_{1}\tilde{\phi}_{2}.
\end{split}
\end{align}
The cost of erasing the $x$ dependence has been to introduce derivatives which should be understood to act solely on the fields $\tilde{\phi}_{1,2}$.

\section{The generating functional in phase space}\label{app.mean_values}
In this Appendix we will briefly review how to compute the generating functional $Z_{DBC}$ with Dirichlet boundary conditions on phase space \cite{Bonezzi:2012vr}. Using the notation of mean values, the definition of the generating functional in terms of arbitrary sources $k(t), j(t)$ is
\begin{align}
\begin{split}
Z_{DBC}[k,j]:&=\left\langle e^{\int_0^1 dt \left(p\, k +q\, j \right)}\right\rangle_{DBC} \\
&=\frac{\int_{DBC}\mathcal{D}P\ e^{-\frac{1}{2}\int_0^{1} dt\,P^t B P
+\int_0^1dt\,P^tK}}{\int_{DBC}\mathcal{D}P\ e^{-\int_0^{1} dt\,P^t B P}}.
\end{split}\end{align}
In this last expression we have defined the vectors in phase space
\begin{eqnarray}
    P:=\left(\begin{array}{c}p(t)\\q(t)\end{array}\right),&
    K:=\left(\begin{array}{c}k(t)\\j(t)\end{array}\right),
\end{eqnarray}
and the matrix valued differential operator operator
\begin{equation}\label{propagator}
B:=
\begin{pmatrix}
2&-i\partial_t\\
i\partial_t&0
\end{pmatrix}.
\end{equation}

We obtain the
generating functional in phase space simply by completing squares and inverting the operator $B$ -- taking into
account the Dirichlet boundary condition $q(0)=q(1)=0$. The result is
\begin{equation}\label{z}
Z_{DBC}[k,j]=e^{\frac{1}{2}\int_0^1 dt\, K^t B^{-1} K}\, ,
\end{equation}
where the kernel of the operator $B^{-1}$ is given by
\begin{equation}
B^{-1}(t,t')=
\begin{pmatrix}
\frac{1}{2}&\frac{i}{2}\left[h(t,t')+f(t,t')\right]\\
\frac{i}{2}\left[h(t,t')-f(t,t')\right]&2g(t,t')
\end{pmatrix}\, ,
\end{equation}
and we have introduced three auxiliary functions
\begin{align}
\begin{split}
h(t,t'):&=1-t-t'\,,\\
f(t,t'):&=t-t'-\epsilon(t-t')\,,\\
g(t,t'):&=t(1-t')H(t'-t)+t'(1-t)H(t-t')\,.
\end{split}.
\end{align}
In these expressions the sign function $\epsilon(\cdot)$ is $\pm 1$ if its argument is
positive or negative, respectively, while  $H(\cdot)$ represents the
Heaviside function.

\section{Additional formulas regarding the four-point function}\label{app.four_point}
The relevant terms in the computation of the  divergent part of the four-point function are obtained by performing a small proper-time expansion in the general result \eqref{eq.master_formula} for $n=2$. These are
\begin{align}\label{eq.four_point}
\begin{split} \Gamma^{(4)}_{1-loop}&=- \frac{\mu^{-\epsilon}}{4!}\frac{\lambda^2}{8}  \int_0^{\infty} \frac{dT}{\left(4\pi \right)^{D/2}}  \frac{ e^{-m^2 T}}{T^{D/2-1}}\int_{-\infty}^{\infty} \frac{dq_{1}}{(2\pi)^D}\frac{dq_{2}}{(2\pi)^D} \psi(-(q_1+q_2))\\
& \hspace{-0cm}\times \left\lbrace \frac{4!}{2}\tilde{\phi}_1\tilde{\phi_2} + \beta \left[ \frac{1}{T} \alpha'_{\mu\nu} \left( \frac{1}{2} \delta^{\mu\nu}-\frac{T}{4} \delta^{\mu\nu}(J,B_{22}^{-1}J)-\frac{T}{4}C[J^{\mu}]C[J^{\nu}]\right)+\gamma'  \right] \right\rbrace .
\end{split}
\end{align}

The following formulas involving integrals of the $B^{-1}$  and the functional $C$ kernels will prove useful in performing the computation of eq. \eqref{eq.four_point}:

\begin{align}
 \begin{split}
 \int_0^1 dt_1 dt_0 \left( B^{-1}_{22}(t_0,t_0)-B^{-1}_{22}(t_0,t_1)\right) &=\frac{1}{6},\\
 \int_0^1 dt_1 dt_0 ( C(t_0)-C(t_1) )^2 &=-\frac{1}{3}.
\end{split}
\end{align}

Additionally, since it is sometimes easier to work with an explicit expression of the  $\phi_{\star}^4$ interaction in terms of the fields and their derivatives,  the following formulas are useful:
\begin{align}\label{eq.phi_fourth}
 \int dx \, \phi\, (\phi \star (\phi \star \phi))&= \int dx \phi^4 +\beta (s_1+s_2)  \int dx \phi^3  \frac{2}{3}  \left( (D+2) +2 x^{\mu} \partial_{\mu} \right) \partial^2\phi.
\end{align}
An explicit expression of $\phi_{\star,(1)}^4$, the linear term in $\beta$ of the quartic interaction under the integral sign, can be read from \eqref{eq.phi_fourth}.

\end{document}